# Cross-stakeholder service orchestration for B5G through capability provisioning

Vilho Räisänen, Mohammed Elbamby, Dmitry Petrov
Nokia Bell Labs
Espoo, Finland
{first.last@nokia-bell-labs.com}
dmitry.a.petrov@nokia-bell-labs.com

*Abstract*—Cross-stakeholder service orchestration is a generalization of 5G network slices which has potential to increase business agility in Beyond 5G (B5G). An architectural framework is proposed which enables domain operators to expose their functionalities towards E2E services as capabilities. Capability orchestration is proposed as a mechanism for exposure. The use of intent-based management for communicating domain owner's business goals to capability orchestration is analyzed. The combination of business goal input and capability orchestration provides a basis for agile monetization of domain resources for domain owners, and a building block for rich end-to-end B5G services.

## I. INTRODUCTION

Provision of digital services is evolving quickly. The advent of 5G networks makes it possible to provide high-availability and secure edge computing with short latencies by means of network slices. This facilitates new kinds of value networks and a greater variety of business models. In moving towards Beyond 5G (B5G), it is important to support business model differentiation for a variety of stakeholders.

The 5G stand-alone (SA) network architecture [1] provides a platform for services by means of network slices supporting both connectivity and processing in Communication Service Provider (CSP) hosted clouds. ETSI Zero-Touch Network and Service Management (ZSM) [2] provides a framework for End-to-end (E2E) service provisioning spanning multiple domains in a multi-vendor operator environment. ETSI ZSM architecture encompasses APIs towards domains and can be viewed as a means of using network resources programmability. At the same time, ZSM is based on separation of concerns and supports domain autonomy. Programmability for RAN domain is supported by Radio Interface Controller (RIC) in ORAN architecture [3]. The Network Exposure Function (NEF) [4] allows for controlled access to information about the network status to applications run in operator's cloud platforms.

In moving to beyond basic SA 5G, certain trends can be identified. 5G network architecture is going to be used also in private networks within enterprises. For this kind of applications, automation of service provisioning is important [5][6][7]. It also reduces Total Cost of Ownership for 5G CSP networks [8]. With the advent of true dynamic slicing, efficient use of 5G network capacity is important. Market-based mechanisms for monetizing network slices is an enabler for this [9]. Edge computing is expected become highly important with the general availability of edge clouds [10][11]. Artificial intelligence has many potential use cases in mobile networks and can be expected to be used for enhancing effectiveness of resource utilization [12].

An analysis of requirements for 6G networks has started, and relevant concepts are outlined in [13]. A part of this analysis is the expected evolution of the roles of stakeholders and emergence of new ones. Private networks and traditional operator networks are expected to cooperate with each one and other stakeholders. An overview of technical aspects relevant to B5G evolution has been provided in [14]. Business aspects of 6G such as decoupling of growth from cost are addressed in [8] A framework for B5G/6G service provisioning based on provisioning of capabilities was proposed in [15] and provides context for this article. Shortly, capabilities provided by business stakeholder are composed together into E2E services. Later in the article, we show an example where a particular control application is exposed as a capability. This approach requires means for stakeholders to flexibly reconfigure their value offering. In this article, we explore the concept of capabilities and how they can enable domain owners to monetize their resources in an agile manner.

The interface for expressing business requirements is important to facilitate agility. Policy transformations are a mechanism for converting high-level goals into low-level parameter changes [16]. An alternative approach is based on knowledge graphs, where high-level concepts can be represented in a knowledge model and mapped to low-level configurations by means of a reasoner [17][18].

In this article, we assume that intent-based management [19] paradigm is used, supporting expression of business goals of a stakeholder. The intents are then converted to relevant technical configurations (e.g., by one of the means listed above), taking into account potential concurrent closed-loop automation actors and other boundary conditions. Presenting automation to user is important for using management based on high-level goals [20]. The intent in our approach is a statement of goal state rather than description of the means of achieving it. As we shall see, intent management can be used for multiple purposes in B5G capability provisioning.

Below, we shall discuss B5G service provisioning architecture supporting E2E service provisioning with capabilities from one or more domain providers. We proceed to discuss network management aspects of the article from the viewpoint of domain operator. Network management is relevant here since it is also assumed to have intent-based interface which affects capability composition. We then move on to discuss capability orchestration itself from business viewpoint, including intent-based control of capability orchestration. Next, E2E service orchestration leveraging capability orchestration is addressed. We illustrate the architecture by means of examples and conclude with a summary.

## II. B5G SERVICE PROVISIONING

6G networks are expected to cover a wide variety of requirements from a variety of stakeholders [13]. Automated support for cross-actor service provisioning is an enabler for business flexibility and efficient monetization of stakeholder resources. The approach we have studied is based on provisioning of services at sub-slice granularity across stakeholders. The value offering of stakeholders is based on capabilities which are made available for E2E service orchestration by other stakeholders [15]. Such interaction can take place within a capability marketplace or be based on a more static interaction such as a repository.

We propose capability provisioning architecture shown in Figure 1. Capability orchestration is applied within a stakeholder and composes functionalities in one or more stakeholder's domains into capabilities which are exposed for End-to-End (E2E) service orchestration. We assume that a capability is associated with a Service Level Agreement (SLA) towards E2E service orchestration. Relevant SLA parameters could be e.g. availability and instantiation / execution time for capabilities which are exposed as Software as a Service (SaaS). When marketplace approach is used, both price and SLA information are assumed to be available. Additionally, metadata about the service interface and the content of the capability need to be exposed.

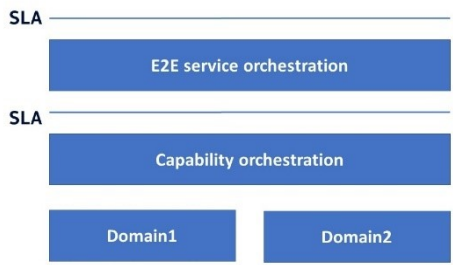

*Figure 1: Architecture for capability provisioning in B5G. Capability orchestration is performed within a business stakeholder and may span multiple resource domains.*

A stakeholder composing an E2E service is thus assumed to have a variety of capabilities available, a subset of which can be used for orchestrating the ultimate service. Aspects of automated service composition have been discussed e.g. in [17]. The E2E service may have an SLA itself, assuming that it is made available to other stakeholders. We shall provide examples of capability and E2E orchestration in Section IV.

In what follows, we shall discuss capability orchestration architecture.

## III. CAPABILITY ORCHESTRATION ARCHITECTURE

In this section, we shall discuss the capability orchestration, starting from network management of domains, proceeding to capability orchestration itself, and concluding with E2E service orchestration.

### A. Network management aspects

Capability orchestration instantiates functionalities in domains. Network management is needed for parameterizing domain functionalities and optimizing them for the current and future capability instances in accordance to business goals of the stakeholder operating the domain. The functionalities of a domain are show in Figure 2. For mobile network, domain orchestration can involve closed-loop automation both for implementing intents as well as for adapting different parts of the network to changes in other parts and the operating environment. For clouds, declarative configuration of Kubernetes [21] is a simple example of a mechanism for intent implementation.

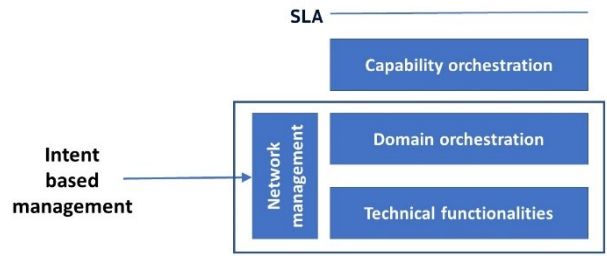

*Figure 2: Functionalities of a domain and intent-based management interface.*

Starting with LTE, closed-loop network automation has been available in the form of Self-Organizing Network (SON) [22]. The simplest implementation is based on fixed rule sets but may be amended with machine learning and keeping track of the effect of the effects of previous configurations [23]. The instances of SON functions need to be coordinated to avoid conflicting configurations. The creation of coordination logic is a complex problem, and the use of policy transformation for this has been studied in (Frenzel, 2016). A method based on the use of automated classical reasoning and supporting ML is outlined in [18][24].

With the above methods, top-level policies or business ontologies can be used to input business requirements to network management of a domain. Irrespective of the machinery used, the target for B5G should be the use of high-level intents which describe the goal state of the system. For example, energy saving intent could be interpreted as switching off small cell layers during nighttime. Closed-loop automation would adapt remaining layers to the effect of intent execution. A concurrent operational Cell Coverage Optimization (CCO) type SON function for a particular cell in scope of the intent might lead to conflict and to issuer of the intent to be informed of infeasibility in the situation at hand. In general, for automation of intent-based management, it is important to keep the human user of the system appraised of the status of automation [20].

*B. Capability orchestration*

Capability orchestration is a key enabler for monetization of domain resources in accordance with business goals of the domain owner. The interfaces of capability orchestration are illustrated in Figure 3 and discussed below.

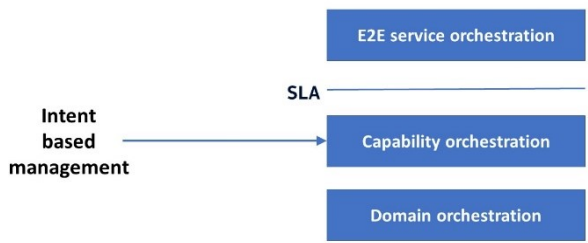

*Figure 3: Interfaces of capability orchestration.*

Interface to service orchestration exposes capabilities for cross-stakeholder E2E orchestration. Metadata describing the capability, input/output parameters, and capability SLA is part of the interface towards the E2E service orchestration. Orchestration interface also supports messages related to creating an instance of the capability.

Interface to domain orchestration facilitates composition of capabilities from functionalities in one or more domains. Consequently, messaging towards a specific domain orchestration may be iterative and dependent on the functionality negotiation with another domain.

Capability orchestration also needs an interface for input of business requirements. An "aggressive" orchestration policy would mean that a large share of domain functionalities would be used for capabilities, which could affect the SLAs provided for individual functionalities. Obviously, the pricing of capabilities could be affected by "aggressive" intents as well. Conversely, a "lax" intent could mean that smaller part of capacity is utilized but provided SLAs could be higher. Various differentiation strategies can be used in hybrid scenarios.

Capability orchestration implements assurance to ensure SLA towards E2E service orchestration and can trigger orchestration of domain resources to maintain target levels.

*C. E2E service orchestration*

The purpose of E2E service orchestration is composition of service with a service SLA guarantee. Service orchestration may be employed by a stakeholder providing services to others, or for stakeholder's own use. Service orchestration makes use of capabilities and their metadata (including capability SLAs).

Service compositions may use same set of capability providers for all service instances or choose the set of capability providers per service instance. The latter case could be useful, for example, for creating local instances of a service which is available over a wide area. In such a case, local capability providers could be leveraged.

The mechanism by which capabilities are identified by service orchestration is beyond the scope of this article.

IV. USE CASE EXAMPLES

In this Section, we illustrate the capability orchestration architecture by means of examples. The first one illustrates selection of capabilities based on technical criteria (in this case latency and mobility). The second one demonstrates orchestration for localized service instances.

*A. Latency-dependent orchestration*

The first use case calls for a wide area control application which requires a certain upper limit for latency towards the cloud platform on which the application is hosted. The "wide area" refers to the need of macro layer mobility. The coordination of autonomous driving for a fleet of vehicles as well as mobility and connectivity could be an example of this kind of application. In this case, the control units in autonomous vehicles would be matched with coordination logic in the cloud.

The latency limit affects the positioning of the coordination logic: for short latency limit, coordination logic needs to be close to the endpoint, which may also mean that the respective control application needs to be handed over to another edge computation platform as the endpoints move. This requires tight coordination between mobile access and computation. For a larger delay limit, the control application may reside further away, and application mobility is not equally critical.

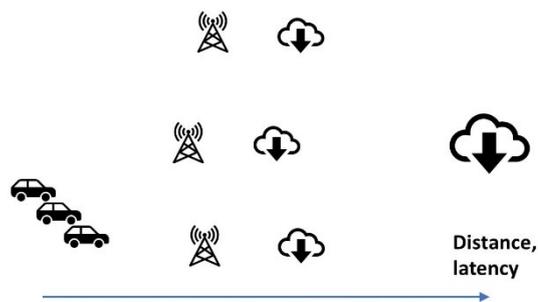

*Figure 4: An illustration of factors relevant to capability orchestration for the first example: latency requirements affects positioning of the control application, which in turn affects the requirement for coordination between radio access and computing platform.*

From the viewpoint of orchestration, a mobile network provider may provide different capability options:

- Only connectivity
- Connectivity + execution platform for coordination application.
- Connectivity + execution platform + mobility

All these options could be simultaneously exposed for E2E service orchestration. A stakeholder composing the coordination service would use information about pricing and capability SLA levels for selecting the best suited option for the service.

The intents expressing business goals for capability orchestration for this example would most likely attempt to maximize revenue stream from different options in view of the momentary availability of resources. Similarly, network management intents would be dependent on the availability of the resources, as well as actual and predicted usage of the different options.

### B. Instance-specific orchestration

In the second use case, we consider orchestration which can create service instances by using local capabilities. The service in question is robotic control, and is provided over a wide area (e.g., a country). It comprises software in robots that are being controlled, control application run on a cloud platform, and connectivity between the two. Additionally, control applications communicate between themselves.

The basic capabilities relevant to this use case are:
- Execution platform for control applications.
- Connectivity between execution platform and robot.
- Connectivity between execution platforms.

In addition, control application may require dedicated AI/ML capabilities from the execution platform, depending on the implementation of robotic control.

In this case, service orchestration knows the effect that needs to be achieved with a combination of capabilities for every instance. Capability providers are selected per instance, taking into account boundary conditions such as latency mentioned in the first use case. In Figure 5, the upper instance supports one robot with a network slice, whereas the second instance supports two robots with one slice, two 6G base stations, and one execution platform.

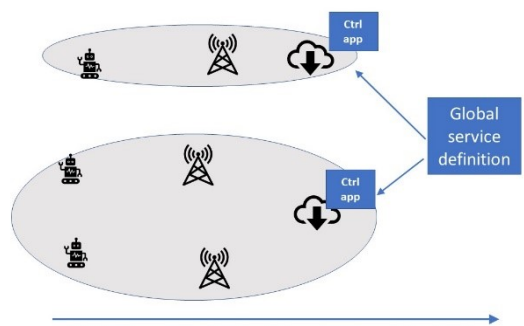

*Figure 5: Local instances for global service definition. A service instance involves instance of control application on an execution platform (blue rectangle) and connectivity (grey ellipse).*

As an example of using business goals in this case, a Digital Service Provider (DSP) type stakeholder could use capability orchestration intent to optimize orchestration for control applications for low latency. Similarly, network management intent would prioritize low latency over high network utilization. This would correspond to the upper case in the Figure.

An alternative approach would be to maximize network utilization, where capability orchestration would attempt to pool instances together while providing a higher end-to-end latency. Similarly, network management intent would prioritize high RAN utilization over strict performance guarantees. This would correspond to the lower case in the Figure.

## V. SUMMARY

We described drivers for service provisioning evolution from 5G towards 6G. There may be novel societal requirements for service provisioning in 6G. We discussed technical enablers such as ETSI ZSM, AI, and edge computing. The value network for service provisioning is expected to become more complex. The roles of current stakeholder are expected to change, and new ones to emerge. The ability to flexibly monetize provider's resources is expected to be important for 5G evolution and especially for the 6G era. The level of complexity for the latter is expected to be greater.

We introduced capability provisioning as an enabler for cross-stakeholder provisioning in B5G. It provides business-agile way of monetizing domain resources for operators and IT providers, and a basis for E2E service marketplace.

We described a high-level architecture for capability provisioning with interfaces towards E2E service management as well as domain orchestration. The architecture was analyzed from the viewpoints of network management and intent-based business requirements interface. We illustrated the architecture by means of a use case.

This article is merely an outline for the work that needs to be carried out in unlocking value in domain providers' systems. For example, further work is needed in analyzing capabilities which 6G DSPs can provide based on their infrastructure.

ACKNOWLEDGMENT